\documentclass{aa}
\usepackage{graphicx}

\def\planck{{\sc planck}}
\def\healpix{{\sc healpix}}
\def\hfi{{\sc hfi}}
\def\lfi{{\sc lfi}}

\def\fknee{{f_{\mathrm{knee}}}}

\def\fspin{{f_{\mathrm{spin}}}}
\def\fmarg{{f_{\mathrm{marg}}}}
\def\frp{{f_{\mathrm{rpt}}}}
\def\f1{{f_{\mathrm{1}}}}
\def\ft{{f_{\mathrm{t}}}}
\def\fr{{f_{\mathrm{r}}}}
\def\fk{{f_{\mathrm{k}}}}

\def\tspin{{t_{\mathrm{spin}}}}
\def\trpix{{t_{\mathrm{rpix}}}}

\def\nrpix{{n_{\mathrm{rpix}}}}
\def\nring{{n_{\mathrm{ring}}}}

\def\nring{{n_{\mathrm{ring}}}}
\def\ndet{{n_{\mathrm{det}}}}

\def\ncross{{n_{\mathrm{cross}}}}

\def\Cdiag{{C_{\mathrm{diag}}}}
\def\Coff{{C_{\mathrm{off}}}}

\def\Fdiag{{F_{\mathrm{diag}}}}
\def\Foff{{F_{\mathrm{off}}}}

\def\Eblock{{E_{\mathrm{block}}}}
\def\Erow{{E_{\mathrm{row}}}}

\def\Grow{{G_{\mathrm{row}}}}

\def\Nr{{\cal N}_{\mathrm{r}}}
\def\Nt{{\cal N}_{\mathrm{t}}}

\def\Ns{{\cal N}_{\mathrm{s}}}

\def\WO{{W_{\mathrm {0}}}}
\def\W1{{W_{\mathrm {1}}}}

\def\mone{{r_{\mathrm {1}}}}
\def\mtwo{{r_{\mathrm {2}}}}
\def\mthree{{r_{\mathrm {3}}}}
\def\mfour{{r_{\mathrm {4}}}}

\def\mT{{r_{\mathrm {T}}}}
\def\mQ{{r_{\mathrm {Q}}}}
\def\mU{{r_{\mathrm {U}}}}

\def\ph#1{{\phi_{\mathrm {#1}}}}
\def\n#1{{n_{\mathrm {#1}}}}

\def\bA#1#2{{{A}_{\mathrm {#1}}^{\mathrm {#2}}}}

\def\Q#1{{Q_{\mathrm {#1}}}}
\def\U#1{{U_{\mathrm {#1}}}}

\def\hQ#1{{\hat{Q}_{\mathrm {#1}}}}
\def\hU#1{{\hat{U}_{\mathrm {#1}}}}

\def\Qoff#1{{\delta^{\mathrm Q}_{\mathrm {#1}}}}
\def\Uoff#1{{\delta^{\mathrm U}_{\mathrm {#1}}}}

\def\alph#1{{\alpha_{\mathrm {#1}}}}
\def\bet#1{{\beta_{\mathrm {#1}}}}

\def\cm#1{{c_{\mathrm {#1}}}}

\def\sm#1{{s_{\mathrm {#1}}}}

\def\dsamp{{\Delta_{\mathrm{s}}}}

\def\offdelta#1{{\delta_{\mathrm{#1}}}}

\def\sigmat{{\sigma_{\mathrm{t}}}}
\def\sigmar{{\sigma_{\mathrm{r}}}}
\def\sigmax{{\sigma_{\mathrm{x}}}}
\def\sigmaoff{{\sigma_{\mathrm{off}}}}

\def\Xoff{{X_{\mathrm{off}}}}

\def\j0{{j_{\mathrm{0}}}}

\def\simlt{\lower.5ex\hbox{$\; \buildrel < \over \sim \;$}}
\def\simgt{\lower.5ex\hbox{$\; \buildrel > \over \sim \;$}}
\def\simggt{\lower.5ex\hbox{$\; \buildrel \gg \over \sim \;$}}

\def\eg{{\rm e.g.}}
\def\ie{{\rm i.e.}}

\def\etal{{\rm et al.}}
\def\eqn{{\rm Eq.}}
\def\eqns{{\rm Eqs.}}
\def\etc{{\rm etc}}
\def\fig{{\rm Fig.}}

\def\dsp{\displaystyle}
\def\l#1{\left#1}
\def\r#1{\right#1}

\def\apj{{\sl Ap.J.}}

\def\aa{{\sl A\&A}}

\def\prd{{\sl Phys.\ Rev.\ D.}}

\begin{document}

\title{The effects of low temporal frequency modes on minimum
variance maps from \planck.}

\author{
Radek Stompor\inst{1}
\and 
Martin White\inst{2}}

\institute{Computational Research Division, Lawrence Berkeley National Laboratory, Berkeley, CA 94720\\
Space Sciences Laboratory, University of California, Berkeley, CA 94720
\and
Departments of Astronomy \&\ Physics, University of California, Berkeley, CA 94720}

\date{\today}

\abstract{
We estimate the effects of low temporal frequency modes in the time stream
on sky maps such as expected from the \planck\/ experiment -- a satellite
mission designed to image the sky in the microwave band.
We perform the computations in a semi-analytic way based on a simple model
of \planck\/ observations, which permits an insight into the structure of
noise correlations of \planck-like maps, without doing exact,
computationally intensive numerical calculations.  We show that, for a
set of plausible scanning strategies, marginalization over temporal
frequency modes with frequencies lower than the spin frequency of the
satellite ($\simeq 1/60$ Hz) causes a nearly negligible
deterioration of a quality of the resulting sky maps.  We point out
that this observation implies that it should be possible to successfully remove 
effects of long-term time domain parasitic signals from the
\planck\/ maps during the data analysis stage.

\keywords{methods: data analysis - statistical;
cosmology: cosmic microwave background}
}
\authorrunning{R.$\;$Stompor \&\ M.$\;$White}
\titlerunning{Low frequency modes and map-making for \planck }

\maketitle

\section{Introduction and motivation.}
\label{sect:intro}

The \planck\/ satellite\footnote{\planck\/ home page:\\
{\centerline{\tt http://www.astro.esa.estec.nl/$\sim$Planck}}}
is designed to measure the microwave sky with unprecedented sensitivity
and angular resolution.
While its primary objective is to characterize the anisotropies in the
Cosmic Microwave Background (CMB), the full sky maps in each of the 9
frequency channels will be the true \planck\/ legacy.

However, making maps from an instrument like \planck\/ is non-trivial,
due to the way the sky is scanned and the long-term correlations in
the noise properties of the instruments (so-called $1/f$ noise).
Though a lot of attention has been given to the long term stability of
the \planck\/ instrument during its design phase it is practically
unavoidable that slow secular drifts and semi-periodic signals
will be present in the actual \planck\/ data (\eg, Seiffert~\etal,
2001, Mennella~\etal, 2002) and will require a software solution.
In fact long term drifts, on time scales of hours and longer, are
conspicuous in the recent data of the WMAP satellite and have been
found to be of sufficient importance that a special treatment was
applied to the time ordered data prior to turning them into the sky
maps (Hinshaw \etal, 2003).  It seems therefore prudent to assume that
some kind of an analogous procedure will be required to minimize the
impact of such effects on the \planck\/ maps.

Here we attempt to estimate the magnitude of this impact and elucidate
the interplay of those parasitic signals with the usual $1/f$ low
frequency tail of the time domain noise as expected for \planck.
More specifically we investigate two (related) questions.
First we strive to gain an understanding of how much the `striping' in
a map is constrained by low-frequency information in the time stream
(and therefore liable to be compromised by the parasitic signal mentioned
above) and how much it is constrained by intersecting scan patterns on the
sky.
We shall find that the presence of $1/f$ noise at any realistic level
means that almost all of our ability to control the low frequency
modes comes from the overlap of scan paths in the spatial domain, rather
than from long time-scale information in the time stream.
Secondly we wish to gain an understanding of how parasitic signals or
specific removal techniques affect the structure of the map error matrix.
Though for definiteness we will focus hereafter on very specific problems,
the analysis presented should be useful in guiding considerations of other
similar issues for \planck\/ as well as a variety of different experimental
setups.

The structure of this paper is as follows.
We start with a quick review of map making techniques
(Section \ref{sec:mapmaking})
and present briefly our assumptions about \planck, its scanning strategy
and anticipated performance (Section \ref{sect:planck}).
In Section \ref{sect:tod}, we analyze the low temporal frequency problem
adopting a time domain perspective, while in Section \ref{sect:pod} we
derive corresponding pixel domain constraints.  These two sections are
quite technical, and the reader may wish to skim these on a first pass.
We compare both results in Section \ref{sect:comp} and finally draw some
conclusions in Section~\ref{sect:sum}.

\section{Map making review} \label{sec:mapmaking}

A reconstruction of a map of the microwave sky from a sequence of
measured time ordered data samples is by now a well understood problem
(\eg, Wright, Hinshaw, \&\ Bennett, 1996, Tegmark, 1997a, Stompor
\etal, 2002).
An entire suite of approaches, ranging from a simple binning of
observations into pixels on the sky, through destriping techniques to
optimal minimum variance map-making, has been to date successfully
implemented, investigated in detail (\eg, Wright \etal, 1996, Tegmark,
1997b, Delabrouille, 1998, Borrill \etal, 2000, Natoli \etal, 2001,
Dor\'e \etal, 2001, Maino~\etal, 2002) and in a handful of cases
converted into publicly available software packages (\eg,
MADCAP\footnote{{\tt http://crd.lbl.gov/$\sim$borrill}},
MAPCUMBA\footnote{{\tt http://ulysse.iap.fr/cmbsoft/mapcumba/}},
MADmap\footnote{{\tt http://crd.lbl.gov/$\sim$cmc}}).
These different approaches trade, to various degrees, simplicity and
numerical speed for accuracy of the produced maps, and usually they
can be derived either as a special case of, or an approximation to, a
more general formalism based on a maximum likelihood approach to
map-making.
The latter provides a handy and concise algebraic formulation of the
entire problem (Tegmark, 1996).
It has been demonstrated within such a formalism that a number of relatively
straightforward generalizations are possible allowing us, in a statistically
strict manner, to account for a variety of systematic problems
commonly troubling actual experimental data (Wright \etal, 1996,
Oliveira-daCosta \etal, 1999, Stompor \etal, 2002, Hinshaw \etal,
2003).
All these developments highlight the practical feasibility,
flexibility and robustness of this approach.  In this context a major
task in a successful accomplishment of the map-making stage for any
CMB experiment becomes the detection and characterization of
systematic contributions plaguing real data sets.  The problem is made
more difficult by the sheer size and complexity of current and future
data sets.  Consequently, for map-making practitioners, the map-making
issue is far from being definitely resolved and keeps providing
challenges on a case-by-case basis.

In a standard map-making procedure (Tegmark 1997a, Borrill \etal,
2000) a final uncertainty of the map is encoded in 
% a single mathematical object -- 
a pixel-pixel noise correlation matrix. In the
case of a Gaussian instrumental noise, this matrix provides a complete
description of the map error. However, in most of circumstances of the
interest, given the size of forthcoming maps, the computation and
storing of such a matrix is prohibitive, let alone any investigation
of its structure and its dependence on a presence of parasitic signals
in the data, or on details of a removal technique applied to such a
systematic.  In such cases an understanding of a role of a particular
systematic effect and its impact on the final map needs to be
investigated by different usually more indirect and simplified means.

\section{{\sc planck} -- basic features.}  \label{sect:planck}

To elucidate the impact of instrumental parasitics on low temporal
frequencies (generally understood hereafter as those lower than the
satellite spin frequency) we construct a simple model of \planck\/
which captures all of the essential features for our purposes.
We assume that the satellite spins on its axis roughly once per minute
($\tspin\equiv {1/\fspin}=1$min), with any given detector beam sweeping
out a circle (or more generally -- following Wandelt \&\ G\'orski, 2001,
and Wandelt \&\ Hansen, 2003 -- a basic scan path) on the sky with opening
angle of $2\theta_0=170^\circ$.
We will refer to a short stretch of a circle (ring) of a beam-size length
as a ring pixel.
The sampling rate is assumed to be
$\dsamp=(\theta_{\mathrm{fwhm}}/3)/(11800 \fspin)$, corresponding to
three measurements per the FWHM of a detector beam ($\equiv \theta_{\mathrm{fwhm}}$)
at each passing. 
Hereafter, whenever needed we will assume that $\theta_{\mathrm{fwhm}}=10'$.
Each circle is observed for $1$hour, before the
satellite axis re-points and a different ring (shifted with respect to
the previous one by $\sim 2.5'$ along the great circle on the sky) is
observed for another time $T$.  We will assume
that the re-pointing from a circle to a next one is always
instantaneous.
As the satellite axis is re-pointed only by $\sim 2.5'$ every hour,
during every four hour long period each detector of the satellite
observes approximately the same sky sweeping during that time a single 
ring of a beam-size width.
Hence, hereafter as basic scan paths we will consider sky rings as swept out during a time period of
$T=4$hours. The re-pointing frequency is thus, $\frp= 1/T \simeq 7\cdot10^{-5}$Hz.  
The number of scans of each ring is ${T/ \tspin}$ and there are $\nrpix$
(beam size) ring pixels per ring.
As we will discuss in some detail in Sect.~\ref{sect:ringnoise},
$\nrpix$ also corresponds to a number of (nearly) uncorrelated, independent
ring pixels for a single ring of a \planck-like experiment and is therefore
an important factor in the scaling relations of the pixel domain
noise properties derived in the following.

We should emphasize that though the discussion presented here touches
upon and depends upon a number of scan-related assumptions, the
issues involved in optimizing the scanning strategy for \planck\/ are
more complex and varied than what is included in the discussion
presented below.  Clearly a more pointed analysis is required to
elaborate on those issues. We leave such a discussion for future
work.  As far as this paper is concerned, while it is clear the
numerical prefactors in our results are dependent on a particular
scanning strategy, we believe that our main conclusions are
independent on their specific details.

Hereafter, we assume that the Gaussian instrumental noise can be
accurately described as, 
\begin{equation}
P\l(\ft\r)=\sigmat^2\,\dsamp\,\l(1+{\fknee\over \ft}\r), 
\label{eqn:noisespec}
\end{equation} 
where $\ft$ is a temporal frequency, $\fknee$ and
$\sigmat^2$ parameterize the spectrum, and, $\sigmat$ describes the (white)
noise level per measurement.
 Both these values are strongly
detector-dependent.  In the following, whenever necessary, we will
assume the following values: for \hfi\/, $\fknee \simeq 30$mHz and
$\sigmat\simeq1000\mu\mathrm{K}\ (\simeq
100\mu\mathrm{K}\sqrt{\mathrm{sec}}/\sqrt{\dsamp})$ and for \lfi,
$\fknee \simeq 50$mHz and $\sigmat\simeq 3000\mu\mathrm{K}\ (\simeq
300\mu\mathrm{K}\sqrt{\mathrm{sec}}/\sqrt{\dsamp})$ (in both cases in
antenna units). Here, \hfi\/ and \lfi\/ refer to the High and Low Frequency Instruments
of \planck\/ respectively. The actual flight parameters may be somewhat better.
However, for any anticipated values of $\fknee$ the following relation is satisfied,
\begin{equation} 
 \frp \ll \fspin \simlt \fknee.
\end{equation}
As it is discussed below, our conclusions will not be affected by changes 
of either the noise level, which will only lead to a rescaling of the estimated values, or
the assumed values of $\fknee$ as long as these are larger than $\simgt 1$mHz.

\section{ Time domain considerations.}  \label{sect:tod}
\subsection{Background}

The generalized least squares map-making starts by modeling the
observation as
\begin{equation}
  t = As + n
\end{equation}
where $t$ indicates the time stream data,
$A$ is the pointing matrix, $s$ the sky signal and $n$ the noise.
If $\Nt=\langle n n^t\rangle$ is the (time-domain) noise correlation matrix,
the minimum variance estimator of the sky signal, $m$, is
\begin{equation}
  m = \left[ A^t \Nt^{-1} A\right]^{-1} A^t \Nt^{-1} t
\label{eqn:optimap}
\end{equation}
with $\left[ A^t \Nt^{-1} A\right]^{-1}$ the pixel-pixel noise correlation
matrix.
This equation can be efficiently solved using iterative methods if
$\Nt$ is known (Dor\'e~\etal, 2001, Natoli~\etal, 2001). Those methods do not require explicit computations of
the pixel-pixel noise correlation matrix -- a property which makes
solving for the map estimate feasible from both CPU time and disk/memory
storage points of view.

Clearly, in this equation two kinds of constraints are incorporated:
first based on the known correlations in the time domain, as given by
the noise correlation matrix, $\Nt$, and second based on the fact that
some parts of the (unchangeable by definition) sky are observed
multiple times during an experiment, as encoded in the pointing
matrix, $A$.  Both constraints are usually tightly combined together,
\eg, as seen in \eqn~\ref{eqn:optimap}, giving us the `best'
possible map estimate, for instance, that with a minimum noise variance.

Efforts to remove/minimize systematic effects present on the time
stream level can affect both of these types of constraints to a
different extent. If, for instance, entire stretches of the time
ordered data stream are rendered unusable and effectively removed from
the data, some pixels on the sky may not be observed multiple times anymore
or even observed at all, clearly affecting the strength of the pixel domain
constraints.
If, in the contrary, only some of the temporal frequency bands have been
found to be compromised by a systematic effect, it will be 
time domain constraints which will be affected more directly.

In the following we will focus on the latter case assuming that only
low temporal frequencies have been contaminated by a systematic effect
of some kind but no time samples have had to be removed and therefore
no changes to an actual scan pattern observed on the sky have been
made.

\subsection{Destriping vs optimal map-making for \planck.}

Given the rather simple scanning strategies planned for \planck, the
sky map recovery can proceed in a number of ways. In particular,
multiple re-scanning of the successive rings on the sky opens up the
possibility of performing map-making in two stages.  In this case,
first maps of the single rings are created which are subsequently
combined together using the fact that different rings cross each other
multiple times on the sky, \ie, exploiting the pixel domain constraint
in a parlance of the previous Section.  Such a two-step approach is a
keystone of the {\sl destriping} methods
(Dellabrouile, 1998, Maino~\etal, 2002, Keihanen~\etal, 2003).
Techniques of this kind are in general not optimal.
That is the case  even if single ring
maps are produced using optimal map-making
(as we will assume in this paper henceforth),
rather than simple sky binning as in `traditional' destriping.
This fact can be understood as follows.  As long as the beam is simply
sweeping the sky in a circle, in the frequency domain the sky signal
is concentrated only at harmonics of the spin period and at zero
frequency. The latter contains not only actual sky monopole but also
all the higher multipole moments which do not average to zero over the
circle. All together, these contributions amount to a constant ring
offset. In a total power CMB experiment the constant offset of a single segment
of the time ordered data is usually lost, leading to a loss of the ring
offsets together with the corresponding part of the actual sky signal.

Let us assume now that after spending time $T$ on a given ring, the
observation continues and the instrument is re-pointed and a scan of
another circle commences, as it is the case for \planck. 
In the destriping methods, the continuity of the time stream from a ring to ring is ignored, and each ring is analyzed separately.
Consequently, as in a single ring case mentioned above,
the offsets of all the rings are lost 
during the initial single-ring processing
and need to be subsequently restored (from now on only relative).
The uncertainty involved in such a procedure
increases the effective noise level in the map and a noisier map results.
The situation looks different in the methods in which the time stream is not cut into 
segments, which are then analyzed separately, but treat the time stream in its entirety.
Indeed, if the
re-pointing is performed without interrupting the time stream
continuity, the sky signal formerly confined to the zero frequency
mode and the spin frequency
harmonics, now partially resides at the harmonics of the re-pointing frequency
$\frp(=1/T)$ and at sidebands of the harmonics of the spin frequency
(\ie, $f=\fspin\pm\frp,\fspin\pm 2\frp, \dots; 2\fspin\pm\frp,
2\fspin\pm 2\frp, \dots,$ \etc) as well.
Though the zero frequency mode is again lost, all the signal confined
to the non-zero frequency bands is in principle accessible,
rendering tighter constraints on the recovered sky map.
If the instrument is successively and smoothly re-pointed in such a way
that the observed sky area progressively increases, more of the multipole
moments of the map are contained in non-zero frequencies.
In the case of a full sky covered in such a way,
only the sky monopole would be lost together with the zero frequency mode.

In the following we address the issue of how much of an actual improvement
over the two step method can be expected from such an approach in a case of
a realistic \planck-like experiment.
Given that, on the ring level, the two-stage method exploits only the pixel
domain constraints, while the full optimal map-making attempts to make use
of both pixel and time domain ones, this question is clearly just a rephrasing
of the problem posed in the previous Section.
At the computational level we can turn this problem into a question of how
the time domain derived constraints on the relative offsets of two rings
compare with those derived from the pixel domain.
Alternately, as uncertainty in the recovery of the rings offsets results in
the presence of strongly correlated linear features in the map aligned with
the scan direction (\ie, {\sl stripes}), we will refer to that problem as
striping.

The attractive advantage of the two-step map-making is its robustness
to low frequency parasitic signal removal.
Indeed, if the parasitic signal is confined to the frequency range
$f\le\fmarg\simlt\fspin$, then high-pass filtering that part of the time
stream which comes from each ring can remove most non-cosmological signals
and at most an offset in the sky signal, causing no extra loss in precision.
By contrast, the loss of the precision of the optimal map-making incurred
due to a removal of the contaminated low frequency modes needs to be
investigated in detail (see Section~\ref{sect:todoff}).

At this point it is worth emphasizing two issues. First, in the case
of \planck, where $\fknee > \fspin$, if the final solution is to be
nearly optimal it may need to account for the presence of $1/f$ noise
within each ring. This suggests that \eqn~\ref{eqn:optimap} be solved
on each ring (see Sect.~\ref{sect:ringnoise}) as the first step of the
destriping method.
Second, in this paper we will use the words filtering and marginalization
over a given frequency band interchangeably, always meaning the latter.
Henceforth, marginalization (or filtering) is understood as a procedure
of ``weighting-out'' unwanted frequency modes from the final map
and the corresponding map noise correlation matrix by setting
the noise level in those modes to infinity (\ie, the weights to zero).
In both cases the relevant operations are usually done in Fourier space.
More details on this approach can be found in the next Section and elsewhere
(\eg, Stompor~\etal,~2002).

Of course, the marginalization over the frequency band is not the only
way of dealing with the parasitic signal. Other options can be viable,
especially if extra information about the origin or character of the
problem is available (\eg, Stompor~\etal~2002). By comparison the
marginalization may look like quite a drastic approach. However, as we
show that in the following, for the low temporal frequency parasitic
signal the marginalization turns out to be as good an approach as any
other and more generally applicable.

\subsection{Constraining offsets.}
\label{sect:todoff}

The questions of interest then are how important the low-frequency
information contained in $\Nt$ is in reducing the the stripiness in
the maps and how it depends on filtering out progressively higher
frequencies. These questions can be rephrased as the following:
suppose we observe two rings during time $2T$.  If there is no sky
signal, how well can we constrain a relative offset, $\delta$, between
the two rings using the data $t$?  The variance in the offset obtained
by generalized least squares is $\left[ A^t \Nt^{-1} A\right]^{-1}$
where now $A=[0,0,\cdots,0,0,1,1,\cdots,1,1]^t$ describes the effects
of the offset on the time stream.  If the noise is stationary with
power spectrum as in \eqn~\ref{eqn:noisespec} then the (inverse)
variance is 
 \begin{equation}
  \sigma_\delta^{-2} = {1\over 4} \sum_k\ {\left|\WO\l(\fk T\r)\right|^2\over
  P(\fk)},
\label{eqn:offtod}
  \end{equation} 
  where the sum is over frequencies
  $\fk=k/2T$ from 0 up to the Nyquist frequency and $\WO/2$ is the Fourier
  transform of the pointing matrix $A$ which is given, up to a phase,
  by 
  \begin{equation}
  \WO\l(x\r) \equiv j_0\l( \pi x\r) % \exp \l( -\iota \pi\; x\r)
  \label{eqn:wind0}
  % W(x=\pi \fk T) = {1\over 2} \j0(x) 
  \end{equation} 
where $\j0(x)\equiv\sin(x)/x$ is the spherical Bessel function of order zero.
If the marginalization is performed the amplitude of the inverse noise power
spectrum is set to zero for all frequencies $f\le \fmarg$ and $\sigma_\delta$
correspondingly increases.  
  \begin{figure} 
  \centerline{
  \includegraphics[scale=0.5,angle=0]{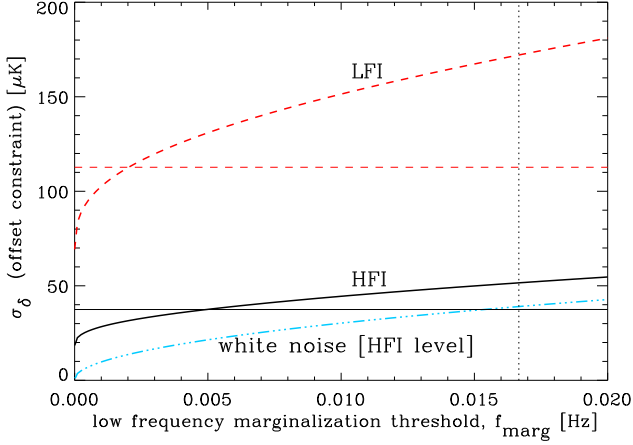} } \caption{ The
  dependence of the time domain constraint on relative offsets as a
  function of the low frequency marginalization threshold
  ($\fmarg$). Three curved lines correspond to the fiducial LFI, HFI
  and HFI-like white noise cases, respectively from top to bottom.
  For comparison, the nearly horizontal lines show the noise per
  beam-size pixel for LFI (top line) and HFI (bottom).  The dotted
  vertical line shows the spin frequency of the \planck\/ satellite.
  } \label{fig:tod} 
  \end{figure}
The rate of that increase depends on the assumed noise power spectrum
parameters as shown in \fig~\ref{fig:tod}.  It is clear that the mere
presence of the $1/f$ component in the spectrum is nearly as harmful
as the marginalization itself.  That is because in such a case the
noise power at frequencies $f\ll \fknee$ is sufficiently high to
effectively ``weight out'' (\ie, to marginalize over) the lowest
frequency tail.  Therefore in the absence of the $1/f$ noise the
losses suffered due to the low-frequency marginalization are the
largest.  By the same token the time domain constraints are the
strongest in a white noise case, though the dependence of
$\sigma_\delta$ on $\fknee$ disappears once $\fmarg \gg \fknee$.  For
values of $\fmarg$ of interest here (\ie, $\fmarg \simlt \fspin 
\ll 1/2\dsamp$) and any $\fknee (\ge 0)$ the following approximate 
formula holds to within few percent, 
\begin{equation} 
\sigma_\delta^2\simeq \sigmat^2
{ \pi^2 \fknee\;\dsamp \over
\log\l(\dsp{ \fknee+\fmarg+\f1}\r) -\log\l( \dsp{ \fmarg+\f1}\r)},
\label{eqn:todscale} 
\end{equation} 
here, $\f1= 1/2T \simeq 3.5\cdot10^{-5}$Hz. This formula
together with \fig~\ref{fig:tod} are the main results of this Section.

Note that in a more realistic case the available time stream data set
will be composed of many ring segments of length $T$.  Therefore,
rather than the single parameter case as in \eqn~\ref{eqn:offtod} we
should consider a case with multiple offsets each assigned to a
different stretch of the time stream and derive the constraints also
for the relative offsets of the non-adjacent segments.  Instead of
deriving the appropriate formulas (which are a straightforward
generalization of the single offset case considered above and follow
the same basic steps as presented in the next Section), we just note
that the relative offset of two segments becomes progressively less
constrained if the time separation of these two segments
increases. That is a consequence of the extra noise power present at
the low frequency end of the spectrum either due to a generic, $1/f$,
noise component or the low frequency marginalization.  Therefore if
there is any gain as a result of incorporating the time domain
constraints on the relative offsets it should be the most significant
on the shortest time scales. Hence we conclude that the formulas
derived above, \eqn~\ref{eqn:offtod} \&\ \eqn~\ref{eqn:todscale},
indeed provide the best case answer to be confronted with the pixel
domain constraints derived later on.

\section{Pixel (ring) domain considerations}

\label{sect:pod}

\subsection{Ring-noise properties.}
\label{sect:ringnoise}
Let $A$ be now a pointing matrix from the time domain to a ring pixel
domain.  Then, as usual, noise correlations in the ring domain can be
expressed as,
\begin{equation}
  \Nr=\l(A^t \Nt^{-1} A\r)^{-1} .
\label{npp}
\end{equation}
With crossing time of a single beam-size pixel denoted as $\trpix$ ($={\tspin/\nrpix}$),
the pointing matrix, for a single ring, is given by:
\begin{equation}
  A\l(t,\ k\r)=1; \ \ \ t - t\;\rm{mod}\; \trpix = k\;\tspin,
\label{atp}
\end{equation}
and is zero otherwise.  Here we number the pixels in a way that ring pixel $0$
is observed at $t=0$.
Given a consecutive and linear pixel numbering along a ring
circumference we have,
\begin{equation}
  A\l(t,\ k'\r)=A\l(t+\l(k'-k\r)\;\trpix,\ k\r),
\label{atp_shift}
\end{equation}
for any two pixels $k$ and $k'$.

Rewriting \eqn(\ref{npp}) in a Fourier domain (and assuming circulancy
of $\Nt$) we have,
\begin{equation}
  \Nr=\l[\l(F A\r)^T \mathrm{diag}\l[{P\l(f_t\r)^{-1}}\r]\l(FA)\r)\r]^{-1}.
\label{npp_fourier}
\end{equation}
here $F$ is a Fourier operator. From \eqn(\ref{atp_shift})
\begin{equation}
  \l(FA\r)_{\mathrm{k}}=\exp{\l( -\iota\, 2\pi \ft k\; \trpix\r)}
  \l(FA\r)_{\mathrm{k=0}}.
\end{equation}
From \eqn(\ref{atp})
\begin{equation}
  \l(FA\r)_{\mathrm{k=0}}= \WO\l( \ft\; \trpix\r)
   \W1\l(\ft\;\tspin,\; T/\tspin\r)
\end{equation}
where
\begin{eqnarray}
%  \WO\l(x\r)&\equiv& j_0\l( 2\pi x\r) \exp \l( -\iota \pi\; x\r)\\
  \W1\l(x,\;n\r)&\equiv& {\sin{\dsp{ (n+1) \pi x }}
  \over \sin{\dsp{\pi x}}} \exp{\l( -\iota\,
\dsp{ n \pi x}\r)},
\end{eqnarray}
and $\WO$ was defined in \eqn~\ref{eqn:wind0}.
Let us assume translational symmetry along the ring, neglecting therefore
``switching on and off'' effects, the noise correlation in pixel domain
can be expressed via ``ring domain'' power spectrum,
${\cal P}\l(\fr\r)$, where $\fr$ is a ring domain frequency expressed in Herz.
The inverse noise spectrum in the ring domain is then given as,
\begin{eqnarray}
  {\cal P}\l(\fr\r)^{-1} &=&
  \sum_{\Delta k,\;\ft}\l|\W1\l( \ft\;\tspin,\; T/\tspin\r)\r|^2
  \l|\WO\l( \ft\;\trpix\r)\r|^2\nonumber\\
  &\times& P\l(\ft\r)^{-1}
  \exp{\l( -\iota 2\pi \l(\ft-\fr\r) \Delta k \;\trpix\r)}\nonumber\\
  &=& \sum_{\ft}\;\l|\W1\l( \ft\;\tspin,\; T/\tspin\r)\r|^2
  \l|\WO\l( \ft\;\trpix\r)\r|^2\nonumber\\
  &\times& % P\l(\ft\r)^{-1}\j0\l(\pi\l(\ft-\fr\r)\trpix\r).
  \W1^{\Re}\l(\pi\l(\ft-\fr\r)\trpix, \nrpix\r)\; P\l(\ft\r)^{-1}.
  \label{eqn:invpring}
\end{eqnarray}
Here, the superscript $\Re$ denotes a real part.
The noise correlation matrix can be computed via a Fourier transform,
\begin{equation}
  \Nr=F^T\; \mathrm{diag}\l[{\cal P}\l(\fr\r)\r]\;F.
\label{npp_final}
\end{equation}
Though this result looks very intuitive, the expression for
${\cal P}\l(\fr\r)$ is less so.
Nevertheless, for the set of parameter values as described
in Sect.~\ref{sect:planck}, the resulting ring-domain power
spectrum turns out to be basically a time-domain spectrum resampled on
a sparser grid of frequencies corresponding to the ring domain.
The corresponding ring domain $\fknee$ is
nearly identical to that in the time domain,
and the resampling ensures that the marginalization
over all frequencies $f\le \fmarg$ does not cause any singularities of
the ring domain noise correlation matrix in addition to that at zero
frequency as long as $\fmarg < \fspin$, \ie, in that case
${\cal P}\l(\fr\r)^{-1} > 0$, if $\fr \ne 0$, and ${\cal P}\l(\fr\r)$ is
well-defined. 
For the beam-size pixels, we find that the off-diagonal noise correlations in
the ring domain are
usually less than a percent of the rms noise value per ring pixel, given by,
\begin{equation}
\sigmar^2 = {1\over \trpix} \sum_{\fr} {\cal P}\l(\fr\r).
\label{eqn:sigmaring}
\end{equation}
In the following, we will
therefore neglect these off-diagonal correlations and consistently
keep assuming that the number of independent, uncorrelated ring 
elements is equal to the number of beam size ring pixels, $\nrpix$.
It is important to point out that the assumption made above does not
automatically force the
noise correlations of a final map to be also diagonal. We
discuss that issue further below.

Having estimated the noise correlation matrix, $\Nr$, we can also
estimate the ring map using \eqn~\ref{eqn:optimap}.  In the following,
in the spirit of the destriping methods we will reconstruct the sky
maps via combining all the precomputed sky rings paying a particular
attention to the noise correlation patterns of the resulting sky maps
and the precision of the recovery of the relative ring offsets.

\subsection{Offsets from ring crossings.}

In this Section we will proceed assuming that there are no off-diagonal
noise correlations within each ring. This assumption has been
justified above. We will show below that this assumption does not
mean that there will be no correlations in the final map, made as a
superposition of all the rings on the sky. We will estimate these
correlations as well as dispersions and compare the results with the
time domain constraints derived earlier.

Yet again, let us define a simple map-making problem, this time the
projection is performed from sky pixel domain, as for
example defined by \healpix\/ (G\'orski~\etal~1998), to the ring domain.
From a set of
all ring pixels for all the sky rings we choose only ring-pixels which cross
with at least
one other ring. We form a vector, $r$, of all estimated temperature values for all such
ring pixels. These are derived
from the time ordered data as described in the previous Section.
We assume that each of the ring pixels included above corresponds to a certain
uniquely defined sky pixel.
Though this may not always be the case, as some ring pixels may span more
than a single sky pixel, it is a useful simplification which we expect to
be broken only occasionally.
We number the sky pixels consecutively counting each
pixel once and form a vector of the corresponding true sky
temperatures denoted, $s$. Consequently, any ring pixel temperature,
$r_{\mathrm i}$, can be modeled as a sky temperature
in a corresponding sky pixel, $s_{\mathrm j}$, plus a ring-pixel instrument
noise, $n_{\mathrm i}$ and 
its own specific offset, $x_{\mathrm k}$. The latter contributes to any
measurement made along a given ring.
Hence we get,
\begin{equation} 
r= A s + n + B x = [A, B] \l[\begin{array}{c}
                                \dsp{s}\\
                                \dsp{x}
                           \end{array}\r]
                 +n.  \label{eqn:offpix} \end{equation}
In the above equation, $A$ is a pointing matrix assigning which sky pixel was
observed at which measurement, and a
matrix, $B$, is deciding to which ring a given ring-pixel belonged and
therefore which offset $x$
is to be added to it.  Under our assumptions, the noise correlation matrix,
$\Nr=\langle n n^t\rangle$,
is diagonal and all of its diagonal elements are identical and denoted as,
$\sigmar^2$.

Our task here is to solve this system of equations to estimate the
noise properties of the recovered ``map'', $m$, which includes estimates of
both the actual sky, $s$, and ring offsets, $x$.
The resulting noise correlation matrix is as usual given as,
\begin{equation} 
\Ns=\l( [ A, B]^t \Nr^{-1} [A, B] \r)^{-1} = \sigmar^2
\l[\begin{array}{c c c}
\dsp{ A^t A} & & \dsp{ A^t B}\\
\dsp{ B^t A} & & \dsp{ B^t B}
\end{array}\r]^{-1}.
\label{nxx}
\end{equation}
Hereafter we will denote the matrix on the right hand side of this
equation with, ${\cal A}$.  This matrix has a block structure with two
diagonal blocks describing the correlations between the sky pixels and the
offsets respectively, the two off-diagonal blocks characterize the
cross-correlations between them.  Of course, not all ring pixels are
involved in this map-making problem explicitly, but only those which
are observed during at least two different ring scans. The
correlations between remaining pixels are equal to the correlations
between offsets of the rings they happen to be intersected by. For
well-cross-linked scanning strategies, where most of the pixels are
observed multiple times, from various directions, the pixel-pixel
correlations need to be computed directly or based on
considerations such as those presented here.
In either case the level of stripiness in the maps can be measured as an
RMS difference between the amplitudes of two rings given by, 
\begin{equation}
  \sigma_\delta=\sqrt{ 2\l(\sigmaoff^2-\Xoff\r)}, 
  \label{eqn:sigdeltapix}
\end{equation}
where $\sigmaoff$ and $\Xoff$ are the variance of the ring offsets recovery
and a cross-correlation between two offsets.

In general calculating $\Ns$ in \eqn~\ref{nxx} is difficult and
commonly requires extensive numerical computations, which easily
become prohibitively expensive.
Instead of doing that here we will consider a set of toy models to gain
insight into the structure of the resulting noise correlations.

\subsubsection{two rings}

Let us begin with a pair of two intersecting rings (or more generally,
basic scan paths). Let the number of crossings (\ie, pixels in common
between the rings) be, $\ncross$. To break a degeneracy due to the
unknown and unrecoverable absolute offset of these rings, we constrain
the offset of the 1st ring to be 0.
Recalling that every crossed pixel is observed twice, the matrix in
\eqn~\ref{nxx} in this case reads,
\begin{equation}
{\cal A}=
\l[\begin{array}{c c}
 \dsp{A^t A,}  & \dsp{ A^t B}\\
 \dsp{ B^t A,} & \dsp{B^t B} 
\end{array}\r]
=
\l[\begin{array}{c c c c c}
 \dsp{ 2}      & \dsp{ 0}      &  \dsp{\cdots} &  \dsp{ 0}      &  \dsp{ 1}       \\
 \dsp{ 0}      & \dsp{ 2}      &  \dsp{\cdots} &  \dsp{ 0}      &  \dsp{ 1}       \\
 \dsp{ \vdots} & \dsp{ \vdots} &  \dsp{\ddots} &  \dsp{ \vdots} &  \dsp{ \vdots}  \\
 \dsp{ 0}      & \dsp{ 0}      &  \dsp{\cdots} &  \dsp{ 2}      &  \dsp{ 1}       \\  
 \dsp{ 1}      & \dsp{ 1}      &  \dsp{\cdots} &  \dsp{ 1}      &  \dsp{ \ncross} \\
\end{array}\r],
\end{equation}
where the last column and row correspond to the offset of the second ring.
The inverse of ${\cal A}$ is readily given by,
\begin{eqnarray}
{\cal A}^{-1}&=&
\l[\begin{array}{c c}
 \dsp{ A^t A,}  & \dsp{ A^t B}\\
 \dsp{ B^t A,} & \dsp{ B^t B}
\end{array}\r]^{-1}
={1\over 2\ncross}\times\nonumber\\
\nonumber\\
&\times&
 \l[\begin{array}{c c c c c}
    \dsp{ \ncross+1} & \dsp{ 1}          &  \dsp{\cdots} &  \dsp{ 1}         &  \dsp{-2}  \\
    \dsp{ 1}         & \dsp{ \ncross+1}  &  \dsp{\cdots} &  \dsp{ 1}         &  \dsp{-2}  \\
    \dsp{ \vdots}    & \dsp{ \vdots}     & \dsp{\ddots}  &  \dsp{ \vdots}    &  \dsp{ \vdots}   \\
    \dsp{ 1}         & \dsp{ 1}          &  \dsp{\cdots} &  \dsp{ \ncross+1} &  \dsp{-2}  \\  
    \dsp{-2}         & \dsp{-2}          &  \dsp{\cdots} &  \dsp{-2}         &  \dsp{ 4} \\
  \end{array}\r].
\end{eqnarray}
The following observations can be now made:
\begin{enumerate}
\item{for a sky pixel, the variance decreases with the number of crossings
very slowly and quickly reaches an asymptotic value of $\sigmar/\sqrt{2}$;}
\item{the off-diagonal terms for sky pixels are non-zero and arise as a result
of solving for the unknown relative offsets between different rings.
These correlations are however suppressed by a factor $\propto \ncross+1$
with respect to the diagonal terms and therefore become progressively less
important as the number of crossings between two basic scan paths increases;}
\item{the error in the determination of the relative offset between two rings
is given by $\sigma_\delta=\sigmar\sqrt{2/\ncross}$;}
\item{the correlations of the offset with the sky pixels increases accordingly
so the off-diagonal term is always a half of the offset variance.}
\end{enumerate}

For two special cases: one where two rings on the sky cross only in two points
and the other when two rings fully overlap, we have,

\noindent $\bullet$\ {\sl two circles -- two crossings ($\ncross=2$):}
\begin{equation}
\begin{array}{ l c l}
\medskip
\dsp{ \sigmax}   & \dsp{ =} & \dsp{ \sigmar {\sqrt{3}\over 2};}\\
\medskip
\dsp{ \sigmaoff} & \dsp{ =} & \dsp{ \sigmar;}\\
\dsp{ \Xoff}     & \dsp{ =} & \dsp{ -{\sigmar^2\over 2}.}
\end{array}
\label{eqn:twocross}
\end{equation}

\noindent $\bullet$\ {\sl two overlapping circles ($\ncross=\nrpix$):}
\begin{equation}
\begin{array}{ l c l} 
\medskip
\dsp{ \sigmax}   & \dsp{ =} & \dsp{{\sigmar\sqrt{\nrpix+1 \over 2\nrpix}}\simeq {\sigmar\over \sqrt{2}};}\\
\medskip
\dsp{ \sigmaoff} & \dsp{ =} & \dsp{ \sigmar \sqrt{ 2\over\nrpix} \ll \sigmax < \sigmar};\\
\dsp{ \Xoff}     & \dsp{ =} & \dsp{ -{\sigmar^2 \over \nrpix} \ll -\sigmar^2}.
\end{array}
\label{eqn:tworings}
\end{equation}
Here, $\nrpix$ is the number of independent ring-pixels along the circumference of each ring and in the case of \planck\/ considered here
it is given as $\nrpix \simeq 2000 \gg 1$, justifying the inequalities and approximations made above. 

\subsubsection{simple ring chains}
\label{sect:ringchains}

Let us focus on circular rings on the sky from now on. We will
assume that those can either cross in two different ring-pixels or
remain disjoint, neglecting therefore -- without any loss of
generality -- cases of two tangent rings.  Yet again let us begin with
a pair of crossing rings. We can add a third ring in a number of
ways. Clearly the most efficient way, from the view point of the
offset constraints, is to add this ring in a way to make it cross each
of the two former rings in two different pixels. The least efficient
way, on the other hand, is when all three rings end up having a total
of two pixels in common. We can keep adding more and more rings
following these two prescriptions.  In the following we analytically
consider, in some detail, the resulting two extreme rings pattern on
the sky.

\noindent $\bullet$
\ {\sl $\nring$ rings all crossing each other in two fixed sky pixels:}\\
A picture for this configuration can be that of scan paths following
great circles on the sky intersecting only at the poles.
Such a scanning strategy cannot be realized in practice for a number of
reasons (\eg, finite angular size of the detector focal plane, finite beam
size) but it is a useful mathematical limit to consider.
Again we force the offset of the first ring to be 0.
The resulting correlations are then given by,
\begin{eqnarray}
{\cal A}^{-1}&=&
\l[\begin{array}{c c}
 \dsp{A^t A,} & \dsp{ A^t B}\\
 \dsp{B^t A,} & \dsp{ B^t B}
\end{array}\r]^{-1}=\\
\nonumber\\
&=&
 \l[\begin{array}{c c r r r r}
    \dsp{ {\nring+1\over 2\nring } }  & \dsp{ {\nring-1\over 2\nring } }     & \dsp{ -{1\over2}}  & \dsp{ -{1\over2}}  & \dsp{\cdots}    &  \dsp{\  -{1\over2}} \\
\\
    \dsp{ {\nring-1\over 2\nring } }  & \dsp{ {\nring+1\over 2\nring }}      & \dsp{ -{1\over2}}  & \dsp{ -{1\over2}}  & \dsp{\cdots}    &  \dsp{\ -{1\over2}}  \\
\\
    \dsp{ -{1\over2}}                 & \dsp{ -{1\over2}}                    & \dsp{ 1}           & \dsp{ {1\over2}}   & \dsp{ \cdots}   &  \dsp{\  {1\over2}}  \\
\\
    \dsp{ -{1\over2}}                 & \dsp{ -{1\over2}}                    & \dsp{ {1\over2}}   & \dsp{  1}          & \dsp{ \vdots\ } &  \dsp{ \vdots\ }     \\
\\
    \dsp{\ \vdots}                    & \dsp{\ \vdots}                       & \dsp{ \vdots\ }    & \dsp{ \cdots}      & \dsp{ \ddots}   &  \dsp{ {1\over2}}    \\
\\
    \dsp{ -{1\over2}}                 & \dsp{ -{1\over2}}                    & \dsp{ {1\over2}}   & \dsp{ {1\over2}}   & \dsp{ \cdots}   &  \dsp{ 1}            \\
  \end{array}\r].\nonumber
\end{eqnarray}
Here the left upper 2-by-2 block describes the correlations between the noise
in the two crossing pixels, and the bottom right
$(\nring-1)$-by-$(\nring-1)$ block shows the correlation and uncertainties in
the recovery of the relative offsets in this case.
Two things are evident:
\begin{enumerate}
\item{The noise per each of the two pixels very quickly reaches an asymptotic
value of $\sigmar/\sqrt{2}$ when the number of rings increases, and at the
same time the two ``antipodal'' pixels become nearly $100$\% correlated
(Janssen~\etal, 1996).}
\item{The offsets variance and cross-correlations do not depend on the number
of rings passing through the two pixels at all and they are equal to their
respective values as in the two-ring case considered before
(\eqn~\ref{eqn:twocross}).}
\end{enumerate}
The stripes in a map produced for this scanning strategy would not go away
with an increasing number of independent rings, as $\sigma_\delta=\sigmar$ 
(Eq.\ref{eqn:sigdeltapix}).

\noindent $\bullet$
\ {\sl $\nring$ rings all crossing each other at two different sky pixels:}\\
In this case a convenient way of visualizing a corresponding ring
assembly is to think of a linear configuration with $\nring$ identical
rings each shifted with respect to a previous one by some small and
constant displacement.  The total number of all ring crossing is
$\ncross=\nring\;(\nring-1)$, and there are $\nring-1$ relative
offsets where, as before, we set the offset of the first ring to 0.
The inverse of the matrix, ${\cal A}$, can be then represented in a block
form as follows,
\begin{eqnarray}
{\cal A}^{-1}&=&
\l[\begin{array}{c c}
 \dsp{ A^t A,} & \dsp{ A^t B}\\
 \dsp{ B^t A,} & \dsp{ B^t B}
\end{array}\r]^{-1}
={1\over 2\nring}
 \l[\begin{array}{c c c}
   \dsp{ C}       &  \dsp{ D^t}    & \dsp{ E^t}\\
   \dsp{ D}       &  \dsp{ F}      & \dsp{ G^t}\\
   \dsp{ E}       &  \dsp{ G}      & \dsp{ H}
  \end{array}\r].
\label{eqn:allcross}
\end{eqnarray}
where the meaning of the blocks is following: $C$ describes the
correlations between the pixels crossing involving ring 0; $F$
contains the correlations of all the other crossings and $D$ defines
the cross-correlations between these two sets of pixels. $H$ describes
the correlations between the offset of all rings, and $E$ and $G$ show
their cross-correlations with corresponding sets of pixel crossings.
Each of these matrices can be explicitly computed as it is shown in
the Appendix. For the main purpose of this paper it is sufficient to notice
that:
\begin{enumerate}
\item{In the limit of many rings ($\nring \simgt 10$) the variance of any
crossed pixel is approximately $\sigmar/\sqrt{2}$, and the
cross-correlations are suppressed by a factor $\propto\nring$.}
\item{The error in the reconstruction of the offsets is
$2\sigmar/\sqrt{ \nring}$ with their correlations equal to
$\sigmar/\nring$
(see \eqns~\ref{eqn:var1} \&\ \ref{eqn:var2}) and both decrease
relative to the pixel noise variance, if the number of rings increases.}
\end{enumerate}
Consequently, we have $\sigma_{\delta}=\sigmar \sqrt{ 2/\nring}$.
It is therefore apparent that this scan pattern not only better constrains
the level of the map stripiness but also that it produces lower pixel-pixel
cross-correlations than the scan considered earlier.

\subsection{Getting (more) real.} \label{sect:real}

The considerations presented above are clearly idealized, though they
capture a number of features of actual sky scans and therefore can
serve as a guidance in understanding the properties and effects of
scanning strategies. The major omission seems to be neglecting the
non-zero width of the actual rings on the sky -- a factor which becomes
increasingly important for scans with a large number of rings.  In
such cases new effects can start playing a role due to additional
overlaps between rings due to crowding.  Therefore in the limit of
large, $\nring$, our conclusions may need amending.

Let us consider the scanning strategy with two fixed crossing points
with a large number of rings.  Partial 
overlap between different
rings at high latitudes becomes unavoidable leading to a quick
improvement in the level of the map stripiness over the numbers quoted
above. How rapid the improvement really is depends on the width of
the rings and their numbers.  For example if we consider rings of
FWHM width and assume that there are no gaps left between the rings
at the equator, we find that it creates
$\sim 2\;\sin^{-1}({1/3})/\pi\; \nrpix \simeq 0.2\; \nrpix$
of extra near crossings, \ie, ring pixel overlaps at which their centers
are displaced relative to each other by less than FWHM/3.
These crossings suppress the relative offsets by an extra factor
$\propto \sqrt{ 0.2\;\nrpix} \simeq 20$,
making this scanning strategy more comparable to the other one from that
point of view.

For the scan with all rings crossing each other in two different
pixels on the sky, there is a limitation on how many rings can be a
part of such a configuration. If we assume again that all the rings have
a width equal to the beam FWHM, an upper limit on the number of rings is
given by the number of beam-size pixels which can be fitted along a
quarter of the circumference of a ring, \ie, $\nring\simlt 5400 \sin
\theta_{\mathrm{0}}/$FWHM. In fact we find that for the \planck\/
scanning pattern with an $85^\circ$ opening angle, this limit can
indeed be achieved within at most a factor of 2.

It is important to note that as we assume that all rings are
uncorrelated, it really does not matter if the rings used for
producing a map were scanned by a single detector or many different
detectors. As long as all the considered rings intersect each other in
two different ring pixels, the upper limit on a number of rings
remains unchanged, determining how well we can do in terms of
constraining the relative ring offsets.  As a result, combining the
data of many such detectors can only improve the noise correlations of
the coadded map inasmuch as the single detector maps were
suboptimal. Moreover, once the limit of a maximal number of
intersecting rings has been reached, adding extra data will not bring any
further gain either in terms of the resulting noise level per pixel or in
terms of a lower relative level of off-diagonal correlations.

By contrast, in the case of multiple detectors which perfectly follow the
same trajectory on the sky, combining the maps produced by each of them
will lower the level of both diagonal and off-diagonal noise correlations
by the same factor ($\propto \ndet$, if the noise properties of the detectors
are identical).  This results in lower noise in the final map but does not
enhance the diagonality of the noise correlation matrix with respect to the
maps made of the data of each of the detector separately.
In particular such a procedure will not suppress ``stripes'', if those are
present in the single detector maps.

\section{Comparison.}
\label{sect:comp}

\subsection{Total intensity maps.}

In Sections 3 \&\ 4 we have derived the limits on the offsets due to
the time and pixel domain constraints respectively.
Here we will compare them for the particular case of a
\planck\/-like experiment.

Let us begin with the time domain constraints.  From \eqn~\ref{eqn:todscale}
(see also \fig~\ref{fig:tod}) for the set of \planck-like values
(Sect.~\ref{sect:planck}) we find that $\sigma_\delta \simeq 20-70\mu$K
(\hfi\/ and \lfi\/ respectively) if no marginalization is applied.
That value becomes $\simeq 50-170\mu$K if the low temporal frequencies all
the way up $\fspin\simeq 16$mHz are marginalized over.
It is apparent that both these values are much too high to facilitate
production of a high quality map of the sky.
Yet the only robust way of improving on these numbers in the time domain is
by decreasing the $1/f$ part of the time domain noise power spectrum.
To get these numbers down to a micro-Kelvin level requires $\fknee$ of the
order of a tenth of mili-Herz (\eqn~\ref{eqn:todscale}),
\ie, two orders of magnitude below the expectation for \planck.

The other way to alleviate the problem is to exploit the pixel domain
constraints.
That can be accomplished by a choice of scanning strategy.
Given the results in Sections~\ref{sect:ringchains} \& \ref{sect:real} we
can estimate the level of stripiness for \planck-like scanning strategies,
\ie, composed of repeatedly scanned rings on the sky.
For our fiducial value of a beam FWHM equal to $10'$ and any of the discussed
scans, we get $\nring \simeq 500$ and $\sigma_\delta \simeq 1.7-5\mu$K.
These numbers are well over an order of magnitude lower than the best
(\ie, with no marginalization) time domain results, demonstrating that the
stripes in \planck-like maps will be predominately constrained by the ring
crossings.  
As discussed earlier a full optimal map-making procedure applied to the time stream data divided into segments of the length 
much longer than $T$ could utilize both types of the information, leading to an improved combined constraint on the ring relative offsets.
However, given our estimates above, the resulting constraints can be tigthened only by at most $1\%$ over the
constraints derived from the pixel domain only.
% Consequently, the marginalization over the long temporal
% modes, relaxing the time domain constraint by a factor $\simeq 2.5$,
% can lead only to an increase of $\simlt 2\%$ of the average level of
% stripes in the final map, and therefore is likely to be irrelevant in practice.
(We have allowed generously here for factors of order of
few possibly missing in our pixel domain analysis, and made use of a
fact that both constraints should be combined in a noise weighted
fashion.)

% The question of whether suppressing the variance of the stripes to the
% few micro-Kelvin level is sufficient will depend on the particular
% application for which the maps are to be used and needs to be answered
% on a case-by-case basis.

That shows that 
the two-step map-making approach is a nearly lossless way of making the sky map from the \planck\/ data.
Moreover, again in the case of the optimal map-making,
the marginalization over the long temporal
modes, relaxing the time domain constraint by a factor $\simeq 2.5$,
can lead only to an increase of $\simlt 2\%$ of the average level of
stripes in the final map.
That demonstrates that for either choice of the map-making approach it is possible to produce the final sky map from \planck, nearly optimal and
free of the parasitic contributions residing at the low temporal frequency end, $\simlt \fspin$.

The question, whether the achievable suppression of the variance of
stripes down to a few micro-Kelvin level, as estimated here, is
sufficient, will depend on a particular application the produced maps
are to be used for and needs to be answered on a case-by-case basis.

\subsection{Polarized maps.}

\label{sect:polmaps}

So far we have focused only on the total intensity maps.
However, some of the results can be generalized to give insight
into polarized map-making with \planck.
This issue was discussed in some detail by Revenu~\etal (2000), here we
briefly highlight some of the related problems.

Let us assume that we have four detectors following each other across the
sky.  Each detector is a total power detector but measures only photons
with a specific polarization as defined by a front mounted polarizer.
For simplicity we will assume that all detectors sample the sky at precisely
the same points along the scan trajectory and that the respective orientation
of the polarizer is different in each of them, but fixed with respect
to the scan direction.
For definiteness in a numerical example below we assume these angles to be
$\ph{1}=0^\circ$, $\ph{2}=45^\circ$, $\ph{3}=90^\circ$ 
and $\ph{4}=135^\circ$, respectively.
As total offsets of any of three Stokes parameters can not be recovered,
we can set them to any arbitrary values and force the offsets of three data
streams to be zero.
The offset of the fourth detector stream is to be determined from the data.
We derive the appropriate formulas in Appendix~\ref{app:singlepol}.
We find that in the limit of many beam-size ring pixels the rms noise per
pixel is twice higher for the $Q$ and $U$ maps than for the $T$ map.
Consequently the respective time domain constraints on the relative
offsets of the two polarized ring maps has to be scaled accordingly,
\begin{equation}
\l[
\begin{array} { l}
\medskip
 {\sigma_\delta^{\l(\mathrm{T}\r)}}\\
\medskip
 {\sigma_\delta^{\l(\mathrm{Q}\r)}}\\ 
 {\sigma_\delta^{\l(\mathrm{U}\r)}}   
\end{array}
\r]
\simeq \sigma_\delta^{\l(\mathrm{0}\r)}
\l[
\begin{array} {c}
\medskip
   \dsp{1}\\
\medskip
   \dsp{\sqrt{2}}\\
   \dsp{\sqrt{2}}
\end{array}
\r]
\label{eqn:poloff}
\end{equation}
where $\sigma_\delta^{\l(\mathrm{0}\r)}$ is given by \eqn~\ref{eqn:offtod}. 
The values in \eqn~\ref{eqn:poloff} are to be compared with the pixel domain
constraint derived for each of the Stokes parameters.

The derivation of the pixel domain constraint requires a conversion of
the recovered Stokes parameters from the ring to global coordinates.
For the latter we will use the usual coordinates where the preferred
direction is set along the meridians on a sphere.
In general such a transformation introduces an extra angle dependence to the
``pointing matrix'' of the map-making problem including the offsets
(\eqn~\ref{eqn:offpix}).  That is because what used to be just an
offset in a ring coordinate system in the global coordinates may vary
from a pixel to a pixel as a different rotation may be needed, to 
perform the transformation from the ring to the global coordinate
system.  As a result the crossings will constrain differently the
relative offsets of two rings, depending on the precise geometry of
the intersection.

Although all of that makes the precise computation rather cumbersome,
for the two examples of the scanning strategies considered here the
general problem simplifies allowing for a fast estimate of the effect.
In the case of scanning along great circles intersecting only at the poles
the ring and global coordinate frames basically coincide and the derivation
of the pixel domain constraints follows from our previous considerations
leading to analogous results
(though with both types of the constraints rescaled as in
\eqn~\ref{eqn:poloff}).

In the case of a scan pattern made of rings intersecting each other in
two points, a conversion from the ring coordinates to global coordinates
is necessary, requiring precise knowledge of a particular scan geometry.
However, it seems natural to expect that the cumulative loss with respect
to the total intensity case should be $\sim \sqrt{2}$ as a result of
averaging sine and cosine functions over different possible angles between
two rings at the crossing point.
In fact this intuition can be made more quantitative and shown to be
essentially correct (see Appendix~\ref{app:multiplepol}).
Hence also in this case we derive the same results, within a factor of
$\sqrt{2}$, as in the total intensity case.

Consequently our major conclusion about the irrelevance of the low
(sub-spin) frequencies for the sky map quality holds also in the
polarization case.  Note that again our statements are comparative and
do not state that the residual level of stripiness is satisfactorily
low, but rather that it is mostly determined by the pixel domain
constraints.

\section{Summary.}
\label{sect:sum}

We have investigated the importance of the low temporal frequency
modes on a quality of the sky maps as anticipated from the \planck\/
satellite. We have shown that, for plausible scanning strategies and
predicted instrumental noise properties, marginalization over frequency
modes lower than the satellite spin frequency does not cause any significant
increase in the stripiness of the resulting optimal maps.
This conclusion is based on a fact that the major constraint on the level of
stripes is due to the overlap of the scan rings on the sky and not due to
the low frequency modes contained in the time stream.
Our results also support the idea that the two-step map-making, where
the maps of the scan rings are first computed from the time ordered data, and only then combined together to produce a map of the sky, is nearly optimal and therefore does
not compromise the quality of the resulting map.
Though we have neglected in our considerations of the scanning strategy
possible complications, such as precession or nutation of the satellite spin
axis, we expect that those are bound only to strengthen our conclusion as
they commonly lead to an increase of a level of scan cross-linking.
Therefore, in the context of \planck\/ the results derived above seem to be
quite general and demonstrate the robustness of the mission design with
respect to the long term systematic effects which may be present in the
time ordered data.

It is important to emphasize, that a major factor behind this result
is the presence of $1/f$ noise with an $\fknee$ frequency of the order
of mili-Herz.
In such a case the resulting high noise power at low frequencies leads
to long term variations on its own, the variance of which exceeds the
constraints imposed on the stripes by the scan cross-linking.
If, however, the low frequency excess power were absent, what
would require suppressing $\fknee$ by two orders of magnitude,
both the time and pixel domain constraints would be important in reducing 
the overall stripiness of the resulting map (Mennella~\etal, 2002).

We have also studied the properties of the noise correlations in pixel
domain.
We have shown that though for the beam-size pixels the noise properties
within each
ring on the sky are close to white noise, the final map composed of many
rings can display a non-negligible level of off-diagonal pixel-pixel
correlations.
We have pointed out that though these can in principle be as high as a half
of the pixel variance, for the two scanning strategies considered here the
off-diagonal elements are typically suppressed by a factor $\sim 400-500$,
with respect to the pixel variance, and are therefore of a similar order as
the pixel-pixel noise correlation within each ring due to the $1/f$ noise.
We therefore conclude that the pixel-pixel noise correlation matrix
for \planck-like maps will be strongly diagonal dominated. Whether this observation justifies
neglecting the off-diagonal terms in a statistical analysis of the \planck\/ maps 
may depend on a particular application.

\begin{acknowledgements}
We acknowledge helpful discussions and comments from the US Planck Data
Analysis Team and thank Charles Lawrence for comments on the manuscript.
R.S. is supported by the NASA COMBAT grant no.$\,$ S-92548-F.
M.W. is supported by the NSF and NASA.
\end{acknowledgements}

\appendix

\section{Estimating pixel-pixel noise correlations.}

\subsection{ Total intensity case with every pair of rings crossing
in two different pixels}
\label{app:temp}
The symmetries in this case are somewhat less apparent than the case
treated in the main text.
First of all, we break the symmetry between rings on the sky by arbitrarily
choosing the ring the offset of which is assumed to be zero.
Also the difference between correlations of a pair of pixels which belong to
the same ring and those which do not needs to be made adding yet another
layer of structure to the matrix.
Nevertheless, a great deal of insight can still be gained without the need
for massive (and in practice often prohibitive) numerical computations.

Hereafter we elucidate block-by-block the structure of the matrix ${\cal A}$
as defined in \eqn~\ref{eqn:allcross}.
Let us start with the diagonal blocks. 
The matrix, $C$ can be decomposed into $2\times2$ block matrices, such that,
\begin{equation}
C=
 \l[\begin{array}{c c c c}
   \dsp{ \Cdiag}     &  \dsp{ \Coff^t}   & \dsp{ \cdots}  & \dsp{ \Coff^t} 
\\
   \dsp{ \Coff}      &  \dsp{ \ddots}    & \dsp{ \vdots}  & \dsp{ \Coff^t}
\\
   \dsp{ \vdots}     &  \dsp{ \cdots}    & \dsp{ \ddots}  & \dsp{ \Coff^t}
\\
   \dsp{ \Coff}      &  \dsp{ \cdots}    & \dsp{ \vdots}  & \dsp{ \Cdiag}
  \end{array}\r];
\end{equation}
here the $i$th diagonal block express the correlations of every two pixels belonging to the ring 0 and
the ring $i$, while the off-diagonal blocks describe the cross-correlations between these pairs of pixels.
Note that all these pixels belong to the ring 0 and hence their variance is somewhat lower than all the other pixels considered below.
The blocks can be written explicitly as follows,
\begin{eqnarray}
\Cdiag&=&
 \l[\begin{array}{c c}
   \dsp{ \nring+1}   &  \dsp{ 1}\\
\\
   \dsp{ 1}          &  \dsp{ \nring+1}   
  \end{array}\r];
\label{eqn:var1}
\\
\nonumber\\
\Coff&=&{1\over 2}
 \l[\begin{array}{c c}
   \dsp{ 1} &  \dsp{ 1}
\\
   \dsp{ 1} &  \dsp{ 1} 
  \end{array}\r].
\end{eqnarray}

The matrix, $F$, encodes the correlation between all pairs of pixels belonging to the same (non-zero) ring with all other (non-zero) rings.
Again, it can be represented as,
\begin{equation}
F=
 \l[\begin{array}{c c c c}
   \dsp{ \Fdiag}     &  \dsp{ \Foff}    & \dsp{ \cdots} & \dsp{ \Foff^t} 
\\
   \dsp{ \Foff}      &  \dsp{ \ddots}   & \dsp{ \vdots} & \dsp{ \vdots}
\\
   \dsp{ \vdots}     &  \dsp{ \cdots}   & \dsp{ \ddots} & \dsp{ \Foff^t}
\\
   \dsp{ \Foff}      &  \dsp{ \cdots}   & \dsp{ \vdots} & \dsp{ \Fdiag}
  \end{array}\r];
\end{equation}
where the diagonal blocks read,
\begin{equation}
\Fdiag=
 \l[\begin{array}{c c}
   \dsp{ \nring+3}    & \dsp{ 3}
\\
   \dsp{ 3}           &  \dsp{ \nring+3}
  \end{array}\r];
\label{eqn:var2}
\end{equation}
and the off-diagonal blocks come in two flavors depending on whether the
two pairs of pixels in question belong to the same ring or not,
\begin{equation}
\Foff=   
 \l[\begin{array}{c c}
   \dsp{ 1}   &  \dsp{ 1}
\\
   \dsp{ 1}   &  \dsp{ 1} 
  \end{array}\r]
\times\l\{ \begin{array} {c}
\medskip
               \dsp{{5\over 2}}   \\ % with the pixels on the same ring ....
               \dsp{ 2}             % with the pixels on different rings ....
           \end{array}
\r.
\end{equation}
The correlation matrix for the offsets ($H$) is given as,
\begin{eqnarray}
H&=&2
 \l[\begin{array}{c c c c}
   \dsp{ 2}        &  \dsp{ 1}        & \dsp{ \cdots} & \dsp{ 1} 
\\
   \dsp{ 1}        &  \dsp{ \ddots}   & \dsp{ \vdots} & \dsp{ 1}
\\
   \dsp{ \vdots}   &  \dsp{ \cdots}   & \dsp{ \ddots} & \dsp{ 1}
\\
   \dsp{ 1}        &  \dsp{ \cdots}   & \dsp{ 1}      & \dsp{ 2}
  \end{array}\r].
\end{eqnarray}
The off-diagonal blocks are somewhat more cumbersome as many more subcases need to be kept track of.
To represent cross-correlations between the rings offsets and the pixels belonging to the 0th ring, let us define
first two vectors,
\begin{equation}
\Erow= \l[-1\ , \  -1\r] \times \l\{\begin{array} {c}
                                         \dsp{ 2}\\
                                         \dsp{ 1}
                                    \end{array}
 \r.
\end{equation}
Now we can represent the matrix, $E$, as a vector of block matrices, each of the size $(\nring-1)\times 2$, \ie,
\begin{equation}
E=
 \l[\begin{array}{c c c}
   \dsp{\Eblock^{1}} & \dsp{\cdots} & \dsp{\Eblock^{\ncross/2}}
  \end{array}\r],
\end{equation}
where each of the blocks is made with $(\nring-1)/2$ rows of the $1$st type above and $(\nring+1)/2$ rows of the $2$nd type.
The indices numbering the blocks in the above equation are just to distinguish different ordering of the basic vectors within
a block.
In a similar manner we can represent the $G$ matrix, but this time the basic vectors need to be defined as,

\begin{equation}
\Grow= \l[-1 \ ,\ -1\r] \times \l\{\begin{array} {c}
                                      \dsp{ 2}\\
                                      \dsp{ 3}
                                   \end{array}
 \r.
\end{equation}

All these calculations are clearly quite mundane and using these
results directly in a case of particular ring arrangement on the sky
may not be necessarily straightforward. However, the results presented
above demonstrate explicitly that one can gain some insight into the
structure of the correlations of \planck-like maps using simplified
semi-analytical considerations.
There are a number of important conclusions, which can be drawn from
these results. Let us first recall that we have separated a factor
$1/2/\nring$ in \eqn~\ref{eqn:allcross}, which now needs to be taken
into account.
Thus we find that for this kind of strategy {\sl all} off-diagonal
terms scale as $1/\nring$ with a coefficient at most of the order of
few. The same turns out to be true for the offsets, the variance of
which, as expressed by the $H$ matrix above, decreases with a number
of rings in the scan in the same way as the level of the off-diagonal
terms. However, the variances of the sky pixels themselves are nearly
independent on the number of the rings for any large $\nring$ and
approximately equal to $\sigmar/\sqrt{\nring}$.

\subsection{Polarized case.}

\subsubsection{Single ring.}
\label{app:singlepol}
Let us first consider a single ring on the sky observed by a four polarized detectors. As in Sect.~\ref{sect:polmaps} we assume that all the detector
polarizers are at fixed angles with respect to the ring, which are $\ph{1}=0^\circ$, $\ph{2}=45^\circ$, $\ph{3}=90^\circ$ and $\ph{4}=135^\circ$ respectively.
We also constrain the offsets of the three time streams to be zero and let the offset of the fourth detector ($\equiv \offdelta{4}$) to be undetermined.
As before we first compute a maximum likelihood ring-map for each of available time streams and then convert those into maps of the three
Stokes parameters in the ring coordinate system. The former step is feasible as we assume that the orientations of the detector polarizers
with respect to the ring direction does not change with time. Consequently all the results of Sect.~\ref{sect:ringnoise} apply here directly.
That latter step is again treated like a map-making procedure, however with the noise in the ring pixel domain assumed now to be diagonal
(Sect.~\ref{sect:ringnoise}). The corresponding pointing matrix needs to relate the measurements of any of the detectors with ring maps
of the Stokes parameters. For a single ring pixel it would therefore read,
\begin{equation}
\l[\begin{array} { c}
   \medskip
     \dsp{\mone} \\
   \medskip
     \dsp{\mtwo} \\
   \medskip
     \dsp{\mthree} \\
     \dsp{\mfour}
   \end{array}
\r]
= {1\over2}
\l[\begin{array}{r r r r}
   \medskip
      \dsp{1}   & \dsp{ \cos{ 2\ph{1}}}  & \dsp{ \sin{ 2\ph{1}}} & \dsp{0} \\
   \medskip
      \dsp{1}   & \dsp{ \cos{ 2\ph{2}}}  & \dsp{ \sin{ 2\ph{2}}} & \dsp{0} \\
   \medskip
      \dsp{1}   & \dsp{ \cos{ 2\ph{3}}}  & \dsp{ \sin{ 2\ph{3}}} & \dsp{0} \\
      \dsp{1}   & \dsp{ \cos{ 2\ph{4}}}  & \dsp{ \sin{ 2\ph{4}}} & \dsp{1}
   \end{array}
\r] 
\l[\begin{array} { c}
   \medskip
     \dsp{\mT} \\
   \medskip
     \dsp{\mQ} \\
   \medskip
     \dsp{\mU} \\
     \dsp{\offdelta{4}}
   \end{array}
\r]
+
\l[\begin{array} { c}
   \medskip
     \dsp{\n{1}} \\
   \medskip
     \dsp{\n{2}} \\
   \medskip
     \dsp{\n{3}} \\
     \dsp{\n{4}}
   \end{array}
\r]
\label{eqn:polmap}
\end{equation}
This equation is straightforwardly generalizable for the case of an arbitrary number of the ring pixels.
For definiteness hereafter we order the ring pixels either in a way that first $\nrpix$ values correspond to the total intensity ring-map, next $\nrpix$
values to the $Q$-map and $U$-map for the Stokes parameters maps or simply concatenate together maps for each detector for the detector maps.
The last ($3\nrpix+1$) value is the relative offset of the fourth time stream.
With all these assumptions, the noise correlation matrix for the Stokes parameters determined in the ring coordinate frame reads,
\begin{equation}
{\cal N}
={\sigmar^2\over \nrpix}
\l[\begin{array}{r r r r}
\medskip
 \dsp{ {\cal A}_{\mathrm{TT}}}        & \dsp{ {\cal A}_{\mathrm{TQ}}}        & \dsp{ {\cal A}_{\mathrm{TU}}}        & \dsp{ {\cal A}_{\mathrm{T\delta}}} \\
\medskip
 \dsp{ {\cal A}_{\mathrm{TQ}}^t}      & \dsp{ {\cal A}_{\mathrm{QQ}}}        & \dsp{ {\cal A}_{\mathrm{QU}}}        & \dsp{ {\cal A}_{\mathrm{Q\delta}}} \\
\medskip
 \dsp{ {\cal A}_{\mathrm{TU}}^t}      & \dsp{ {\cal A}_{\mathrm{QU}}^t}      & \dsp{ {\cal A}_{\mathrm{UU}}}        & \dsp{ {\cal A}_{\mathrm{U\delta}}}    \\
 \dsp{ {\cal A}_{\mathrm{T\delta}}^t} & \dsp{ {\cal A}_{\mathrm{Q\delta}}^t} & \dsp{ {\cal A}_{\mathrm{U\delta}}^t} & \dsp{ {\cal A}_{\mathrm{\delta\delta}}} \\
\end{array}\r],
\label{eqn:polnoise}
\end{equation}
where the blocks are given by,
\begin{equation}
{\cal A}_{\mathrm{TT}}=
 \l[\begin{array}{ c c c c}
\medskip
    \dsp{ \nrpix+1} & \dsp{ \; 1}    & \dsp{\cdots} &  \dsp{ \; 1}     \\
\medskip
    \dsp{ 1}        & \dsp{ \ddots}  & \dsp{\vdots} &  \dsp{ \vdots}   \\
\medskip
    \dsp{ \vdots}   & \dsp{ \cdots}  & \dsp{\ddots} &  \dsp{ 1}        \\
    \dsp{ 1}        & \dsp{ \cdots}  & \dsp{1}      &  \dsp{ \nrpix+1}
  \end{array}\r],
\label{eqn:polttpart}
\end{equation}

\begin{equation}
{\cal A}_{\mathrm{TU}}=
 \l[\begin{array}{ c c c c}
\medskip
    \dsp{ -2}       & \dsp{ -2}      & \dsp{\cdots} &  \dsp{ -2}      \\
\medskip
    \dsp{ -2}       & \dsp{ \ddots}  & \dsp{\vdots} &  \dsp{ \vdots}  \\
\medskip
    \dsp{ \vdots}   & \dsp{ \cdots}  & \dsp{\ddots} &  \dsp{ -2}      \\
    \dsp{ -2}       & \dsp{ \cdots}  & \dsp{ -2}    &  \dsp{ -2}
  \end{array}\r],
\label{eqn:poltupart}
\end{equation}

\begin{equation}
{\cal A}_{\mathrm{QQ}}=
 \l[\begin{array}{ c c c c}
\medskip
    \dsp{ 2\nrpix}       & \dsp{ 0}       & \dsp{\cdots} & \dsp{ 0}       \\
\medskip
    \dsp{ 0}       & \dsp{ \ddots}  & \dsp{\vdots} & \dsp{ \vdots}  \\
\medskip
    \dsp{ \vdots}  & \dsp{ \cdots}  & \dsp{\ddots} & \dsp{ 0}       \\
    \dsp{ 0}       & \dsp{ \cdots}  & \dsp{ 0}     & \dsp{ 2\nrpix}
  \end{array}\r],
\label{eqn:polqqpart}
\end{equation}

\begin{equation}
{\cal A}_{\mathrm{UU}}=
 \l[\begin{array}{ c c c c}
\medskip
    \dsp{2\nrpix+4} & \dsp{ \; 2}    & \dsp{\cdots} &  \dsp{ \; 2}    \\
\medskip
    \dsp{ 2}        & \dsp{ \ddots}  & \dsp{\vdots} &  \dsp{ \vdots}  \\
\medskip
    \dsp{ \vdots}   & \dsp{ \cdots}  & \dsp{\ddots} &  \dsp{ 2}       \\
    \dsp{ 2}        & \dsp{ \cdots}  & \dsp{2}      &  \dsp{ 2\nrpix+4}
  \end{array}\r],
\label{eqn:poluupart}
\end{equation}

\begin{equation}
{\cal A}_{\mathrm{T\delta}}=
 \l[\begin{array}{ c}
\medskip
    \dsp{-2} \\
\medskip
    \dsp{\; \vdots} \\
    \dsp{-2}
    \end{array}\r],
\ \ 
{\cal A}_{\mathrm{U\delta}}=
 \l[\begin{array}{ c}
\medskip
    \dsp{-4} \\
\medskip
    \dsp{\; \vdots} \\
    \dsp{-4}
    \end{array}\r],
\ \ 
{\cal A}_{\mathrm{\delta\delta}}=\l[\; 4\; \r].
\label{eqn:poloffpart}
\end{equation}
And all the remaining blocks are equal to zero.
As already remarked in Sect.~\ref{sect:polmaps} the noise matrix is strongly diagonally dominated in the limit of large $\nrpix$.
Note also that the lack of symmetry between the $Q$ and $U$ Stokes parameters, conspicuous in the above equations, is solely due to our choice of which
detector offset is to be determined as a result of the procedure. In fact in the presented here case the fourth detector stream does not contribute to a
constraint on the $Q$ component, as can be seen by directly inverting the relation in \eqn~\ref{eqn:polmap} for our choice of the polarimeters' directions
(Revenu~\etal~2000). Consequently, the unconstrained global offset of the fourth detector map, $\offdelta{4}$, does not cause any extra
cross-correlations of the $Q$-map with itself or any other Stokes parameters. 
In fact the symmetry is restored in the limit of a large number of ring pixels as in this limit the cross-correlation 
due to that offset uncertainty ($\simeq 2\sigmat /\sqrt{ \nrpix}$) becomes irrelevant.

\subsubsection{Multiple rings.}
\label{app:multiplepol}
Let us assume henceforth that the number of ring-pixels is large and, consequently, the off-diagonal terms in
\eqns~(\ref{eqn:polnoise}-\ref{eqn:poloffpart}) are negligible. In this Section, we therefore consider a set of uncorrelated rings of the
three Stokes parameters each defined in a corresponding ring coordinate frame. The offsets of each ring maps are undetermined. As in
Appendix~\ref{app:temp} we assume that each pair of rings intersects in two independent ring pixels.
Hereafter, we focus on the $Q$ and $U$ rings only as spatial rotations do not couple temperature to polarization. The discussion of the
temperature case has been already given in Appendix~\ref{app:temp}, below we show how one can gain some intuition about the results in the
polarized case.
We define a projection from the ring domain to the sky pixel domain as in \eqn~\ref{eqn:offpix}. The ring pixel map is then expressed as,
\begin{equation}
r = \l[A, B\r] \l[\begin{array}{c}
                  \dsp{s}\\
                  \dsp{x}
                  \end{array}\r]
+n,
\label{eqn:polpointsplit}
\end{equation}
where, as before, only intersecting ring pixels are included. This time, however, both the ring pixel map, $r$, and the sky pixel map
$s$, combine together maps of two Stokes parameters, $Q$ and $U$. For future convenience, we define them as follows,
\begin{eqnarray}
r&\equiv&
\l[\begin{array} { c}
   \medskip
     \dsp{\Q{12}} \\
   \medskip
     \dsp{\U{12}} \\
   \medskip
     \dsp{\Q{21}} \\
   \medskip
     \dsp{\U{21}} \\
   \medskip
     \dsp{\Q{13}} \\
   \medskip
     \dsp{\U{13}} \\
   \medskip
     \dsp{ \vdots}\\
   \medskip
     \dsp{\Q{\nring-1, \nring}} \\
   \medskip
     \dsp{\U{\nring-1, \nring}} \\
   \medskip
     \dsp{\Q{\nring, \nring-1}} \\
   \medskip
     \dsp{\U{\nring, \nring-1}}
   \end{array}
\r]
\mbox{\ \ \ }
s\equiv
\l[\begin{array} { c}
   \medskip
     \dsp{\hQ{12}} \\
   \medskip
     \dsp{\hU{12}} \\
   \medskip
     \dsp{\hQ{13}} \\
   \medskip
     \dsp{\hU{13}} \\
   \medskip
     \dsp{\vdots}\\
   \medskip
     \dsp{\hQ{\nring-1, \nring}} \\
   \medskip
     \dsp{\hU{\nring-1, \nring}} \\
   \medskip
     \dsp{\Qoff{1}}\\
   \medskip
     \dsp{\Uoff{1}}\\
   \medskip
     \dsp{\vdots}\\
   \medskip
     \dsp{\Qoff{\nring}}\\
   \medskip
     \dsp{\Uoff{\nring}}\\
   \end{array}
\r],
\end{eqnarray}
where, for instance, $\Q{ij}$ denotes the value of the $Q$ Stokes parameter measured at the intersection of rings $i$ and $j$ during the scan
of the $i$th ring and therefore expressed in the coordinate system of the ring $i$; $\hQ{ij}$ is then a value of $Q$ component in the global
coordinate system, and $\Qoff{i}$ is a $Q$ offset of the $i$th ring. In this convention, $(i,j)$ and $(j,i)$ correspond to two different
pixels where the $i$th and $j$th ring intersect.
Note that we have $\nring$ relative offsets and ($\ncross+\ncross$) values of
$Q$ and $U$ in sky pixel domain to be determined using $4\, \ncross$ measurements. The noise, $n$, is assumed to be uncorrelated and
its variance given by $ \sigmar\sqrt{2}$ (see \eqn~\ref{eqn:polqqpart} \&~\ref{eqn:poluupart}).
With these definitions at hand, the sub-matrix $A$ is then block diagonal,
\begin{equation}
A\equiv
\l[\begin{array}{c c c c}
   \medskip
      \dsp{ \bA{1}{2}}& \dsp{ 0}      & \dsp{ \cdots}  & \dsp{ 0} \\
   \medskip
      \dsp{ 0}       & \dsp{ \bA{1}{3}} & \dsp{ \vdots}& \dsp{ \vdots} \\
   \medskip
      \dsp{ \vdots}  & \dsp{ \cdots} & \dsp{ \ddots}  & \dsp{ \vdots} \\
      \dsp{ 0}       & \dsp{ \cdots} & \dsp{ \cdots}  & \dsp{ \bA{\nring-1}{\nring}}
   \end{array}
\r],
\label{eqn:polring}
\end{equation}
with 4-by-4 blocks given as,
\begin{equation}
\bA{i}{j}
= \l[
\begin{array} { r r r}
\medskip
   \dsp{ \cos{ 2\alph{ij}}}  &  & \dsp{ \sin{ 2\alph{ij}}} \\
\medskip
   \dsp{ -\sin{ 2\alph{ij}}}  &  & \dsp{ \cos{ 2\alph{ij}}} \\
\medskip
   \dsp{ \cos{ 2\bet{ij}}}  &  & \dsp{ \sin{ 2\bet{ij}}} \\
\medskip
   \dsp{ -\sin{ 2\bet{ij}}}  &  & \dsp{ \cos{ 2\bet{ij}}}
\end{array}\r].
\label{eqn:ablocks}
\end{equation}
Here $\alph{ij}$ (or $\bet{ij}$) is an angle between the ring coordinate system of the $i$th ring and the global coordinate system at the sky
pixel denoted $(i,j)$ ($(j,i)$ respectively). Hence, $\bA{i}{j}$ performs two rotations from the global to the $i$th ring coordinate system
at the two crossing points of the $i$th and $j$th rings.

The elements of the block $B$ are either 0 or 1, and an element
$B_{\mathrm{i j}}$ is 1 only if the particular crossing corresponding to the $i$th row of the $A$ matrix is between a ring $j$ and some other ring, and
was measured during the scan of the $j$th ring.

In principle, having determined the structure of both matrices, $A$ and $B$, we can calculate the inverse (up to a factor) of the noise correlation matrix in
sky pixel domain, ${\cal A}$, as defined in \eqn~\ref{nxx}.
However, in practice for a general scan pattern, considered here, the computation of ${\cal A}$ turns out to be quite mundane and requires
more specific assumptions.
Instead, as an example, we calculate only the diagonal block of ${\cal A}$ -- one
corresponding to the correlation matrix for the offsets,
and therefore being of the most interest for the consideration contained in this paper.
To do that first note that the corresponding block of the ${\cal A}$ matrix can be computed by partition and reads (\eqn~\ref{nxx}),
\begin{equation}
{\cal A}_{\delta \delta}=\l[ B^t B - B^t A \l(A^t A\r)^{-1} A^t B \r]^{-1}.
\label{eqn:offblock}
\end{equation}
From \eqn~\ref{eqn:ablocks} we have,
\begin{equation}
A^t A=2 I_{\mathrm{\ncross\times\ncross}}
\end{equation}
and also
\begin{equation}
B^t B=\nring I_{\mathrm{\nring\times\nring}}
\end{equation}
from the definition of $B$ in \eqn~\ref{eqn:polpointsplit}. Here, $I_{\mathrm{\n\times\n}}$ stands for an $n$-by-$n$ unit matrix.
Also from \eqn~\ref{eqn:ablocks}, we find that $A A^t$ is block-diagonal with the $(i, j)$ 4-by-4 block given as,
\begin{equation}
\l[A A^t\r]_{\mathrm{i,j}}=
\l[\begin{array}{ c c c c }
      \dsp{ 1} & \dsp{0}  & \dsp{\cm{ij}} & \dsp{ \sm{ij}} \\
      \dsp{0}  & \dsp{ 1} & \dsp{\sm{ji}} & \dsp{ \cm{ji}} \\
      \dsp{ \cm{ij}} & \dsp{ \sm{ji}}     & \dsp{1}  & \dsp{ 0} \\
      \dsp{ \sm{ij}} & \dsp{ \cm{ji}}     & \dsp{ 0} & \dsp{ 1}
\end{array}\r]
\end{equation}
where,
\begin{eqnarray}
\cm{ij}&\equiv& \cos{ 2\l(\alph{ij}-\bet{ij}\r)}\\
\sm{ij}&\equiv& \sin{ 2\l(\alph{ij}-\bet{ij}\r)}.
\end{eqnarray}
These formulas allow us to perform the matrix operations on the right hand side of \eqn~\ref{eqn:offblock} rendering,
\begin{equation}
{\cal A}_{\delta \delta}^{-1}=\l[\begin{array} {c c c c}
   \dsp{ \nring-4}        & \dsp{ 4\,\cm{12}} & \dsp{\cdots} & \dsp{ 4\,\cm{1 \nring}}\\
   \dsp{ 4\,\cm{12}}      & \dsp{ \nring-4}   & \dsp{\vdots} & \dsp{\vdots} \\
   \dsp{ \vdots}          & \dsp{\cdots}      & \dsp{\ddots} & \dsp{ \vdots} \\
   \dsp{ 4\,\cm{1\nring}} & \dsp{\cdots}      & \dsp{\cdots} & \dsp{ \nring-4}
   \end{array}
\r].
\label{eqn:polofffinal}
\end{equation}
In the case of many rings, \ie, $\nring \gg 4 \ge |4\,\cm{ij}|$, the matrix on the rhs of the last equation is nearly diagonal and
we finally derive the approximate expression on the variance of the relative ring offsets in the case of the polarized maps. This reads,
\begin{equation}
\sigma_\delta \sim \sigmaoff \simeq \sigmar \sqrt{{2\over \nring}}.
\end{equation}

\end{document}